\documentclass[twocolumn,showpacs,prb,superscriptaddress]{revtex4-1}

\usepackage{graphicx}
\usepackage{color}
\usepackage{tabularx}
\usepackage{epsfig}
\usepackage{amsmath}
\usepackage{amssymb}
\usepackage{graphicx}
\usepackage{dcolumn}
\usepackage{bm}
\usepackage{wasysym}

\definecolor{grn}{rgb}{0,0,0.54}

\definecolor{bleu}{rgb}{0.,0.,1.}

\newcommand{\bra}[1]{\ensuremath{\langle #1 |}}
\newcommand{\ket}[1]{\ensuremath{| #1 \rangle}}
\newcommand{\expect}[1]{\langle \rangle}

\begin{document}


\title{Valence bond entanglement entropy of frustrated spin chains}

\author{Fabien Alet}
\affiliation{Laboratoire de Physique Th{\' e}orique, Universit{\' e} de Toulouse, UPS (IRSAMC), F-31062
Toulouse, France}
\affiliation{CNRS, LPT (IRSAMC), F-31062
Toulouse, France}

\author{Ian P. McCulloch}
\affiliation{School of Physical Sciences, The University of Queensland, Brisbane, QLD 4072, Australia}

\author{Sylvain Capponi}
\affiliation{Laboratoire de Physique Th{\' e}orique, Universit{\' e} de Toulouse, UPS (IRSAMC), F-31062
Toulouse, France}
\affiliation{CNRS, LPT (IRSAMC), F-31062
Toulouse, France}

\author{Matthieu Mambrini}
\affiliation{Laboratoire de Physique Th{\' e}orique, Universit{\' e} de Toulouse, UPS (IRSAMC), F-31062
Toulouse, France}
\affiliation{CNRS, LPT (IRSAMC), F-31062
Toulouse, France}

\date{\today}

\begin{abstract}
  We extend the definition of the recently introduced valence bond entanglement entropy to arbitrary SU(2) wave functions of $S=1/2$ spin systems. Thanks to a reformulation of this entanglement measure in terms of a projection, we are able to compute it with various numerical techniques for \emph{frustrated} spin models. 
  We provide extensive numerical data for the one-dimensional
  $J_1-J_2$ spin chain where we are able to locate the quantum
  phase transition by using the scaling of this entropy with the block
  size. We also systematically compare with the scaling of the von Neumann entanglement entropy. We finally underline that the valence-bond entropy definition
  does depend on the choice of bipartition so that, for frustrated
  models, a ``good'' bipartition should be chosen, for instance
  according to the Marshall sign.

\end{abstract}

\maketitle


\section{Introduction}
\label{sec:intro}

Entanglement is a fundamental notion of quantum mechanics, that has over the recent years gained popularity as a way to provide new insights in the quantum many-body problem. From
the condensed matter point of view, one of the most interesting
promises of the study of entanglement properties is the possibility to
automatically detect the nature of quantum phases and of quantum phase
transitions.  In this approach, there is no need to provide {\it a
  priori} physical information or input, such as the specification of
an order parameter. The detection can occur through the study of the
scaling (with system size) of various entanglement estimators. For
instance, the scaling of the von Neumann entanglement entropy for
one-dimensional systems is different for critical and gapped systems -
allowing their distinction. For a recent review of various properties
of entanglement entropy in condenser matter, see Ref.~\onlinecite{review}.

To quantify the entanglement between two parts $\Omega$ and
$\bar{\Omega}$ of a quantum system described by a wave-function
$\ket{\Psi}$, one usually invokes the von Neumann entanglement entropy
(vN EE) defined as:
$$S^\mathrm{vN} (\Omega)=-\mathrm{Tr}_{\Omega} \left( \hat{\rho}_\Omega \ln (\hat{\rho}_\Omega) \right),$$
where $\hat{\rho}_\Omega={\rm Tr}_{\bar{\Omega}} \ket{\Psi} \bra{\Psi}$ is the
reduced density matrix obtained by tracing out the degrees of freedom
in $\bar{\Omega}$. Considering that $\Omega$ is a connected region of space
(such as a block of sites in a lattice model), on general grounds one
expects that $S^\mathrm{vN}$ scales not as the volume of $\Omega$, but
rather as the interface between $\Omega$ and $\bar{\Omega}$. This ``area law''~\cite{area} is due to the fact that the value of
$S^\mathrm{vN}$ is independent of whether degrees of freedom in
$\Omega$ or $\bar{\Omega}$ have been first traced out:
$S^\mathrm{vN}(\Omega)=S^\mathrm{vN}(\bar{\Omega})$ and therefore the
size dependence must come from the boundary between the two parts of
the system.

This general area law has been shown to be fulfilled for many physical
wave-functions. However it is also known to be violated for several
examples, where in most cases some type of long-range correlations
develop between degrees of freedom in $\Omega$ and $\bar{\Omega}$. The
most documented situation is the case of one-dimensional (1d) quantum
critical systems where several important results can be derived from
Conformal Field Theory (CFT). For 1d quantum critical wave-functions
displaying conformal invariance, it was shown~\cite{Calabrese} that
$S^\mathrm{vN}=\frac{c}{3} \ln (x)$ where $x$ is the length of the
block of sites $\Omega$. Here $c$ is the central charge of the
corresponding CFT, and periodic boundary conditions are assumed. The
1d area law (where $S^\mathrm{vN}$ saturates to a constant for large
$x$) is fulfilled on the other hand as soon as the correlation length
is finite.

Even though one expects the area law to be valid in most cases, the
situation of corrections to the area law is still unclear in higher
dimensions. CFT predictions are valid only in a few isolated
situations~\cite{Fradkin,Stephan}. Some exact results are available
for a few specific models: for instance, the ground-state of free
bosonic models fulfills strictly the area law~\cite{bosons1,bosons2}
whereas free fermions can display multiplicative log
corrections~\cite{fermions,bosons2}. On the other hand, calculations
for interacting models become rapidly untractable. Numerical
simulations are also not eased by the higher dimensionality. Exact
diagonalization techniques have access to EE, but they are limited to
very small samples sizes which do not allow a test of the area law,
and deviations thereof, in high dimension. EE comes for free within
the DMRG method~\cite{White}, but it is limited to 1d and quasi-1d
systems. Stochastic methods, such as Quantum Monte Carlo (QMC), have
no problem with simulations of systems in higher dimensions, but
unfortunately the vN EE is a quantity that is extremely complex to
measure within Monte Carlo methods (see however recent
progresses~\cite{Buividovich,Caraglio}).

Alternatively, it is possible to define for certain quantum spin
systems, a different quantifier of entanglement through the use of the
Valence Bond (VB) representation. The key idea is that for two quantum
spins 1/2 at sites $i$ and $j$, the singlet state (or VB)
$\ket{\Psi}=\frac{1}{\sqrt{2}}(\ket{\! \! \uparrow_i \downarrow_j} -
\ket{\downarrow_i \uparrow_j} )$ is maximally entangled. It can
therefore be used as a natural unit of entanglement (in the quantum
information community, the entanglement is often measured in units of
singlets). Consider now a VB state where an even number N of spins
$1/2$ are coupled pairwise in singlets, and divide the spins in two
arbitrary sets $\Omega$ and $\bar{\Omega}$. It is simple to see that
the von Neumann entropy $S^\mathrm{vN}$ is equal to the number of VBs
shared between $\Omega$ and $\bar{\Omega}$ (i.e. where one of the two
spins is located in $\Omega$ and the other in $\bar{\Omega}$), times
the constant $\ln(2)$. This constant is just the von Neumann
entropy of a single spin in a VB. In other words, every singlet
that crosses the boundary between $\Omega$ and $\bar{\Omega}$
contributes $\ln(2)$ to the von Neumann entropy. The picture is very
appealing as it provides a simple geometrical interpretation of
entanglement, a quantity which is not always easy to grasp
intuitively. This argument of course is only exact for the case of VB
states, which are simple factorized states. However, it can be shown,
and we will describe this in detail below, that the picture holds for
all singlet states. This is an important property as most
antiferromagnetic systems have a singlet finite size ground-state.

In general, the resulting Valence Bond Entanglement Entropy (VB
EE)~\cite{Alet,Chhajlany} is {\it different} from the von Neumann
entropy, except for the case of VB states where they coincide. There
are however several points of interest for this alternative
description: (i) it fulfills all desired properties of an
entropy~\cite{Alet}, (ii) the VB EE can be easily computed through QMC
methods in the VB basis~\cite{Sandvik}, offering the possibility to
study $d>1$ systems, (iii) the scaling properties of the VB EE display
several interesting features.

Concerning the scaling (with system size) of the VB EE, it has been
shown first numerically and then analytically that for critical
systems in 1d, the VB EE scales logarithmically with block size
$S^\mathrm{VB} \propto \gamma \ln (x)$ (here and in the following the
symbol $\propto$ denotes proportionality up to a constant). On the
other hand, $S^\mathrm{VB}$ converges to a constant for gapped
systems. The same scaling behavior is displayed by the vN EE. The
only difference comes from the prefactor of the log divergence: while
the numerical estimation of $\gamma$ was first reported~\cite{Alet} to
be consistent with $c/3$ (as for vN EE), the analytical results of
Ref.~\onlinecite{Jacobsen} indicate that the two quantities are
different for Heisenberg spin chains $\gamma = 4 \ln(2) / \pi^2 \neq
1/3$, even though the numerical value of $\gamma \simeq 0.279 $ is very close to
$1/3$ ($c=1$ for the Heisenberg chain). We will comment on the
numerical validation of the exact value $\gamma = 4 \ln(2) / \pi^2$ at
a later stage of this paper.

In two dimensions, the situation seems slightly different. For gapped spins systems, the VB EE was shown~\cite{Alet,Chhajlany} to fulfill a strict area law $S^\mathrm{VB}\propto  x$ (with $x$ the linear size of the boundary between $\Omega$ and $\bar{\Omega}$). The same scaling is expected on general grounds for $S^\mathrm{vN}$. In the case of the ground-state of the 2d Heisenberg model displaying N\'eel order with gapless excitations, the VB EE displays a multiplicative logarithmic correction to the area law~\cite{Alet,Chhajlany,Kallin} : $S^\mathrm{VB}\propto  x\ln(x)$. There is no equivalent calculation (analytical or numerical) for the vN EE of the 2d Heisenberg model, for the reasons described above. However, recent DMRG calculations~\cite{Kallin} on $N$-leg ladders and QMC computations of the Renyi entropy of the 2d Heisenberg model~\cite{Hastings}, suggest that the vN EE displays no such multiplicative logarithmic and that the N\'eel ground-state fulfills strictly an area law $S^\mathrm{vN} \propto x$. This can be seen negatively as the scaling of the VB EE does not match the one of the vN EE, showing its limits to discuss the adherence of  the vN EE to the area law in higher dimensions. One should note however that the VB EE is able to distinguish between a gapless and a gapful state through its scaling -whereas the vN EE cannot-, one of the main original and practical motivations of studying entanglement in condensed-matter systems. As a side remark, we also note that the VB EE can be used to characterize shared information in the different context of stationary states of stochastic models~\cite{Stoch}.

In this paper, we investigate the properties of the VB EE using a
different approach.The original VB EE definition\cite{Alet} is
intimately related to the fact it is possible to consistently
define\cite{Mambrini} a VB occupation number able to quantify the
presence/absence of a SU(2) dimer on a given bond for any singlet
state. Note that the definition only depends on the chosen bipartition
of the lattice into two subsets, in spite of the VB basis
overcompleteness. We derive in Sec.~\ref{sec:def} an alternative but
equivalent definition of the VB occupation number which is free of any
VB basis formulation.  This allows to define the VB EE in the
practical $S_z$ basis, and its computation through different numerical
schemes (such as Exact Diagonalization or DMRG) than the VB QMC method
used in previous works. Being now able to compute the VB EE for
frustrated systems, we study in Sec.~\ref{sec:results} both vN and VB
EE for the $J_1$-$J_2$ spin chain, using DMRG techniques. We discuss
and compare how the scaling of both entropies can detect the critical
phase (for small $J_2$) and the quantum critical point that separates
it from the dimerized gapped phase present at large $J_2$. We also
discuss the importance of the Marshall sign present in the
ground-state wave function when comparing the two entropies. We finish
with a discussion in Sec.~\ref{sec:disc} on the usefulness of the
approach, as well as on the further possibilities open by the $S_z$
representation of VB occupation numbers.


\section{VB free formulation of the VB EE}
\label{sec:def}

\subsection{VB occupation number as a projection}

{\it Original formulation --- } In this paragraph we recall some definitions and results on VB occupation number\cite{Mambrini}. Choosing a bipartition of the $N$-site lattice into two equal sized subsets $\cal A$ and $\cal B$, the bipartite VB subspace is generated by all the bipartite VB states
\begin{equation}
\label{eq:bip_rvb}
\vert \varphi_{\cal D} \rangle = \bigotimes_{\substack{(i,j) \in {\cal D} \\ i \in {\cal A}, j \in {\cal B}}} [i,j],
\end{equation}
where $[i,j]$ is a SU(2) dimer state and ${\cal D}$ is a dimer covering of the system. For any bond $(i,j)$ such as $i\in {\cal A}$ and $j\in \cal B$ the VB occupation number in the state $\vert \varphi_{\cal D} \rangle$ is defined as:
\begin{equation}
\label{eq:bond_occupation_pure}
n_{(i,j)}(\vert \varphi_{\cal D} \rangle)=
	\begin{cases}
	1& \text{if $(i,j)$ belongs to $\cal D$},\\
	0& \text{otherwise}.
\end{cases}
\end{equation}
The bipartite VB manifold is overcomplete: all bipartite VB states are singlet ($S=0$) states but their number $(N/2)!$ is much larger than the singlet subspace dimension\cite{Rumer,Mambrini}. As a consequence, a given linear combination of bipartite VB states $\vert \Psi \rangle = \sum_{{\cal D}} \lambda_{\cal D} \vert \varphi_{\cal D} \rangle$ can be rewritten in many alternative linear combinations $\vert \Psi \rangle = \sum_{{\cal D}} \mu_{\cal D} \vert \varphi_{\cal D} \rangle$ with $\lambda_{\cal D} \neq \mu_{\cal D}$. This point requires to reconsider the extension of Eq.~(\ref{eq:bond_occupation_pure}) for linear combinations of bipartite VB states since the identity $\sum_{{\cal D}} \lambda_{\cal D} n_{(i,j)} \left ( \vert \varphi_{\cal D} \rangle \right ) = \sum_{{\cal D}} \mu_{\cal D} n_{(i,j)} \left ( \vert \varphi_{\cal D} \rangle \right )$ is not obviously granted. It is nevertheless possible to prove that $n_{(i,j)}$ is linear\cite{Mambrini} in $\vert \varphi_{\cal D} \rangle$ which provides an intrinsic definition of $n_{(i,j)}(\vert \Psi \rangle)$ despite the bipartite VB manifold overcompleteness.

{\it Projection (VB states) ---} We give here an alternative but equivalent definition of the VB occupation number which (i) is explicitly independent of the way the state is decomposed in the overcomplete bipartite VB basis, (ii) is valid for any spin $S$ and (iii) will be shown to be more versatile for numerical computations. The spin-$S$ dimer is defined as the two-site singlet state:
\begin{equation}
\label{eq:dimerS}
[i,j]_S = \frac{1}{\sqrt{2S+1}} \sum_{s_z=-S}^{+S} (-1)^{S-s_z} \vert -s_z,+s_z\rangle
\end{equation}
We define the reference state
\begin{equation}
\label{eq:reference_state}
\vert R_S \rangle = \vert -S , +S , \ldots , -S , +S \rangle
\end{equation}
where the state is written in the $\otimes_i \hat{S}_z$ eigenstates basis and ordered such as ${\cal A}$ and ${\cal B}$ sites appear in alternating order. In particular $\vert R_{1/2} \rangle$ is nothing but the N\'eel state. As already noticed\cite{Sandvik} for $S=1/2$, the reference state has an equal overlap with all bipartite VB states: $\langle R_S \vert \varphi_{\cal D}\rangle$ does not depend on the bipartite dimer covering $\cal D$. This property is established by a direct evaluation from Eq.~(\ref{eq:dimerS}) and Eq.~(\ref{eq:reference_state}) showing $\langle R_S \vert \varphi_{\cal D}\rangle= 1/(2S+1)^{N/4}$.

For any bipartite VB state $\vert \varphi_{\cal D} \rangle$ and for any bond $(i,j)$ such as $i \in {\cal A}$ and $j \in {\cal B}$ we are going to show that
\begin{equation}
\label{eq:projection_identity}
n_{(i,j)}(\vert \varphi_{\cal D} \rangle)= - \frac{(2S+1)^{N/4}}{2S} \langle R_S \vert \hat{S}^{+}_i \hat{S}^{-}_j \vert \varphi_{\cal D} \rangle .
\end{equation}
Indeed, we have
\begin{equation}
\label{eq:SpSm}
\hat{S}^{+}_i \hat{S}^{-}_j \vert R_S \rangle = (2S) \vert  \ldots , \overbrace{-S+1}^{i}, \overbrace{S-1}^{j} , \ldots \rangle.
\end{equation}
If the bond $(i,j)$ is occupied, $\vert \varphi_{\cal D} \rangle = \ldots \otimes [i,j]_S \otimes \ldots$ and a simple inspection of Eq.~(\ref{eq:dimerS}) shows that $\langle R_S \vert \hat{S}^{+}_i \hat{S}^{-}_j \vert \varphi_{\cal D} \rangle=-(2S)/(2S+1)^{N/4}$ and hence $n_{(i,j)}(\vert \varphi_{\cal D} \rangle)$ as defined in Eq.~(\ref{eq:projection_identity}) is 1. On the other hand, $\vert \varphi_{\cal D} \rangle = \ldots \otimes [i,k]_S \otimes [l,j]_S\ldots$ if the bond $(i,j)$ is unoccupied. The total $S_z$ component on any occupied bond of a VB state is 0. Thus any eigenstate of $\hat{S}^z_i + \hat{S}^z_k$ (or $\hat{S}^z_l + \hat{S}^z_j$) with a non-zero eigenvalue is then orthogonal to $\vert \varphi_{\cal D} \rangle$. It is salient from Eq.~(\ref{eq:SpSm}) that $\left ( \hat{S}^z_i + \hat{S}^z_k \right ) \hat{S}^{+}_i \hat{S}^{-}_j \vert R_S \rangle = +\hat{S}^{+}_i \hat{S}^{-}_j \vert R_S \rangle$ and $\left ( \hat{S}^z_l + \hat{S}^z_j \right ) \hat{S}^{+}_i \hat{S}^{-}_j \vert R_S \rangle = - \hat{S}^{+}_i \hat{S}^{-}_j \vert R_S \rangle$. Hence $n_{(i,j)}(\vert \varphi_{\cal D} \rangle)$ as defined in Eq.~(\ref{eq:projection_identity}) is 0 in this case.

Finally, it is easy to see that if both $i$ and $j$ sites are located in the same subset ${\cal A}$ or ${\cal B}$, the definition Eq.~(\ref{eq:projection_identity}) also ensures that $n_{(i,j)}=0$ which is always true (independently of $\vert \varphi_{\cal D} \rangle$) as no dimer is allowed on such a non-bipartite bond.

Let us mention some of the advantages of definition Eq.~(\ref{eq:projection_identity}) as an alternative to  Eq.~(\ref{eq:bond_occupation_pure}). First of all, it is explicitly linear in $\vert \varphi_{\cal D} \rangle$  which ensures that its extension to arbitrary linear combination of bipartite VB states  can be consistently defined. As stated before, in the case of $S=1/2$, the subspace of bipartite VB states is a basis of the total singlet sector, ensuring that a dimer occupation number can be defined for any $S=1/2$ singlet. This is not true anymore for general spin $S$: bipartite spin-S VB states do not form a basis of spin-$S$ singlets. However it can be shown that they form a basis of the subset of spin-S singlets that are also SU($N$) singlets~\cite{Beach} (with $N=2S+1$) so that Eq.~(\ref{eq:projection_identity}) can be useful in that context.

{\it Projection (VB states superpositions) ---} 
Using Eq.~(\ref{eq:projection_identity}), the occupation number for an arbitrary linear combination of bipartite VB states $\vert \Psi \rangle = \sum_{{\cal D}} \lambda_{\cal D} \vert \varphi_{\cal D} \rangle$ is defined as,
\begin{equation}
\label{eq:projection_mixed}
n_{(i,j)}(\vert \Psi \rangle)= - {\cal N} \frac{(2S+1)^{N/4}}{2S} \langle R_S \vert \hat{S}^{+}_i \hat{S}^{-}_j \vert \Psi \rangle,
\end{equation}
where ${\cal N}$ is a normalization constant. It would be tempting to take ${\cal N}=1/\sqrt{\langle \Psi \vert \Psi \rangle}$.
However $n_{(i,j)}$ is designed to measure the number of dimers (0 or 1) on bond $(i,j)$ and for any VB state a given site $i\in {\cal A}$ is dimerized with another site on sublattice ${\cal B}$. Hence the normalization condition writes,
\begin{equation}
\label{eq:norm_condition}
\sum_{j \in {\cal B}} n_{(i,j)}(\vert \Psi \rangle)= 1.
\end{equation}
This condition enforces,
\begin{equation}
\label{eq:norm}
{\cal N}=\frac{1}{\sum_{{\cal D}} \lambda_{\cal D}} = \frac{1}{(2S+1)^{N/4}\langle R_S \vert  \Psi \rangle},
\end{equation}
and
\begin{equation}
\label{eq:projection_mixed_normalized}
n_{(i,j)}(\vert \Psi \rangle)= - \frac{1}{2S} \frac{\langle R_S \vert \hat{S}^{+}_i \hat{S}^{-}_j \vert \Psi \rangle}{\langle R_S  \vert \Psi \rangle}.
\end{equation}

Contrary to Eq.~\eqref{eq:bond_occupation_pure}, this last expression is {\em explicitly independent} of the linear combination chosen to expand $\vert \Psi \rangle$ on the overcomplete bipartite VB manifold as it only involves projections of $\vert \Psi \rangle$. Since Eq.~(\ref{eq:projection_mixed}) does not give any prominent role to the bipartite VB basis to express $\vert \Psi \rangle$, it will allow numerical computations outside the VB QMC scheme such as with Exact Diagonalization and DMRG (see Sec.~\ref{sec:num}).
 
Note that this expression is potentially singular if $\vert R_S \rangle$ is orthogonal to $\vert \Psi \rangle$. As an example, let us consider a one-dimensional spin-$1/2$ chain with $N=4p$ sites (where $p$ is an integer). If we denote $\cal S$ the spin inversion symmetry $S^z \rightarrow -S^z$ and $\cal T$ the translation symmetry, any $q=\pi$ singlet state $\vert \Psi \rangle$ will transform as ${\cal S} \vert \Psi \rangle = \vert \Psi \rangle$ and ${\cal T} \vert \Psi \rangle = -\vert \Psi \rangle$. Consequently, ${\cal S} {\cal T} \vert \Psi \rangle = -\vert \Psi \rangle$. On the other hand, if the bipartition ${\cal ABAB} \ldots$ is chosen, the reference state given in Eq. \eqref{eq:reference_state} is obviously invariant under ${\cal S} {\cal T}$. As a consequence, $\vert \Psi \rangle$ and $\vert R_S \rangle$ are orthogonal and Eq. \eqref{eq:projection_mixed_normalized} can not be used. 

This issue, which is a direct consequence of the normalization defined by Eq.~\eqref{eq:norm}, suggests that the bipartition and hence $\vert R_S \rangle$ (see Eq.~\ref{eq:reference_state}) may not be chosen regardless of $\vert \Psi \rangle$. More generally, normalizing a state by the sum of its coefficients in an expansion like in Eq.~\eqref{eq:norm}, requires a careful inspection of its nodal structure or equivalently of its Marshall sign which in turn dictates an appropriate bipartition for the reference state. We will further discuss this issue in Sec.~\ref{sec:marshall}.

{\it VB EE ---} Using Eq.~(\ref{eq:projection_mixed_normalized}), the VB EE measuring the entanglement between $\Omega$ or $\bar{\Omega}$ in a state $\vert \Psi \rangle$ can be expressed as,
\begin{equation}
\label{eq:VBEE_proj}
S^{\text{VB}}_{\Omega} (\vert \Psi \rangle)= \ln 2 \times \sum_{{\substack{(i,j) {\text{ such as}} \\ i \in {\Omega}, j \in {\bar{\Omega}}}}} n_{(i,j)}(\vert \Psi \rangle),
\end{equation}
where the spatial sums run over all possible locations of VBs, that is
over all sites $i$ in $\Omega$ and all sites $j$ in $
\bar{\Omega}$. Since we know that $n_{(i,j)}=0$ whenever $i$ and $j$
belong to the same subset ${\cal A}$ or ${\cal B}$, we can restrict
the summation only to the non-vanishing cases.

\begin{figure}
  \begin{center}
   \includegraphics*[width=\columnwidth]{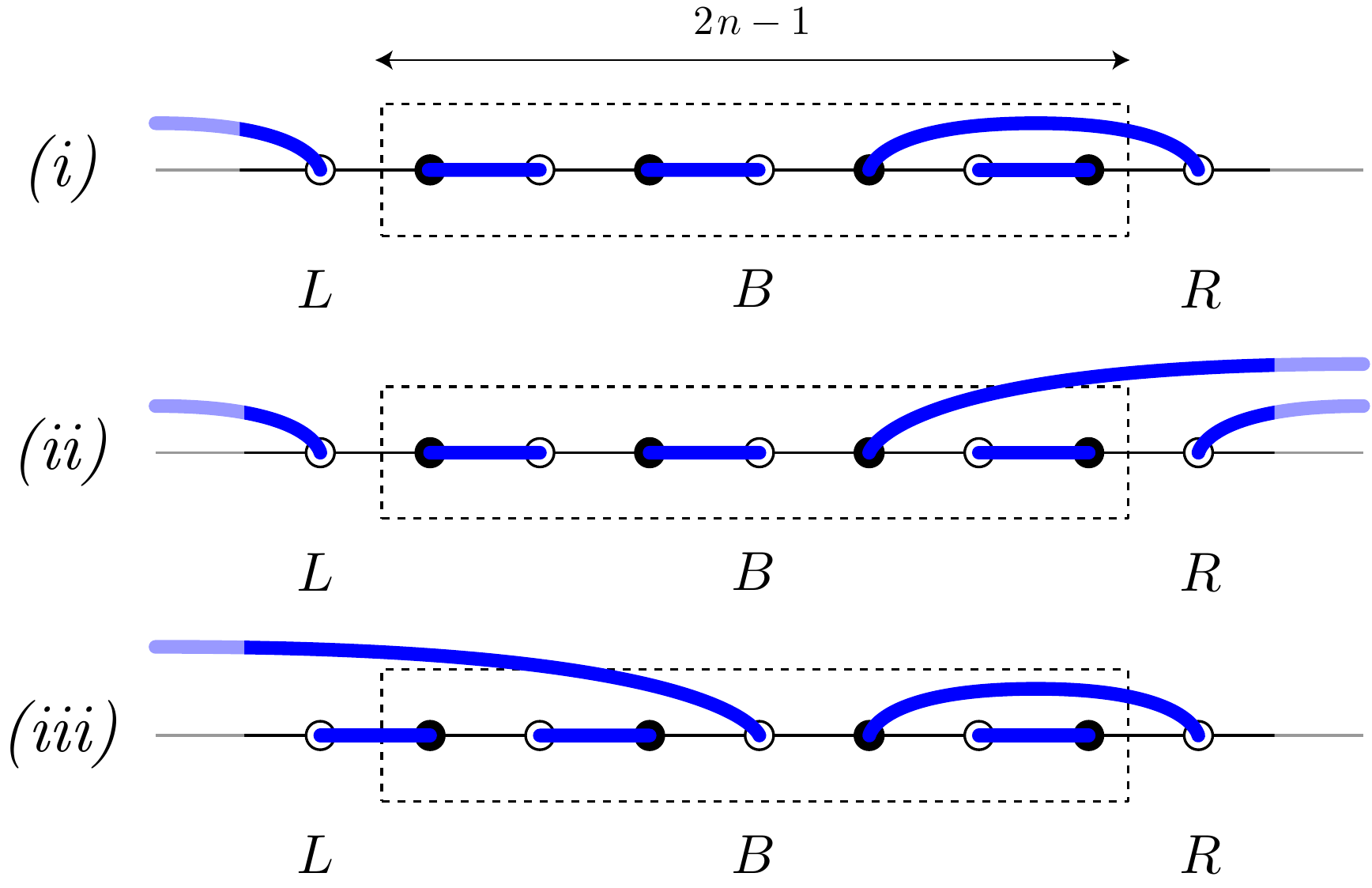}
   \end{center}
  \caption{(Color online) Three possible cases for valence bond configurations in the non-crossing basis for a given block $B$ with an odd number of sites and neighboring sites $L$ and $R$ (see text for details).}
  \label{fig:evenodd}
\end{figure}
{\it VB EE for one-dimensional periodic systems --- } 
We finally make a remark on the behavior of $S^{\text{VB}}_{\Omega}$ for translation-invariant one-dimensional systems.
When $\Omega$ is a linear segment of size $2n$ (with $n$ integer) we find that 
\begin{equation}
\label{eq:evenodd}
S^{\text{VB}}(2n)=\frac{1}{2} \left[ S^{\text{VB}}(2n-1)+S^{\text{VB}}(2n+1) \right],
\end{equation}
which means graphically that $S^{\text{VB}}$ is made of linear
segments. Let us propose an easy graphical proof of this
statement. Consider a VB configuration and let us compare VB EE for
different blocks obtained by adding two extra sites $R$ and $L$ at
each end of a $(2n-1)$-sites block $B$ (see
Fig.~\ref{fig:evenodd}). Since the VB EE is a well defined quantity,
we can choose to work in the complete non-crossing basis~\cite{Rumer}
where VB do not cross according to some one-dimensional ordering of
the sites. Since the $2n-1$ block has an odd number of sites, it is
clear that $R$ and $L$ \emph{cannot} be connected by a singlet. Then,
we can consider all possible cases: (i) $R$ is connected to $B$ but
not $L$, or vice-versa; (ii) neither $R$ nor $L$ are connected to $B$;
(iii) both $R$ and $L$ are connected with $B$. These cases are shown
in Fig.~\ref{fig:evenodd} and from the figure, it is straightforward
to check that $S^{\text{VB}}(B)+S^{\text{VB}}(B+R+L) =
S^{\text{VB}}(B+R)+S^{\text{VB}}(B+L)$. As a conclusion, if periodic
boundary conditions are used so that the entropy only depends on the
number of sites of the block, we deduce Eq.~(\ref{eq:evenodd}).

\subsection{Numerical computations}

From now on, we focus on the case of $S=1/2$ systems.

\label{sec:num}
{\it With Exact Diagonalization --- }
We use the Lanczos algorithm in order to compute the ground-state of
large 1d chains~\cite{Ref_Lanczos} . We also implement lattice
translations as well as fixing the total $S^z$ quantum number in order
to reduce the Hilbert space size so that we can solve systems up to
$N=32$ sites. Once the wavefunction is obtained in the symmetrized
basis, one can easily compute its overlap with the reference state,
which gives the denominator of
Eq.~(\ref{eq:projection_mixed_normalized}). In order to compute its
numerator, we need to apply the operator $ \hat{S}^{+}_i
\hat{S}^{-}_j$ for all pairs of sites $(i,j)$ with $i$ in the selected
block and $j$ outside (let us remind that we can restrict ourselves to the case where $i$ and $j$ belong to different subsets). 
In this case, it turns out that  it is simpler to apply this spin 
operator on the reference state,  since it reduces to a swap operator for
spins 1/2.  Finally, in order to compute the VB EE, and since we are
using translation symmetry, we have to make an average over all the
positions of the block on the chain.

{\it With DMRG --- }
The calculation of the VB entanglement entropy is straightforward, utilizing matrix product techniques to
calculate the overlap between the groundstate and the reference state  $N=\langle \uparrow \downarrow \uparrow \ldots \downarrow \! \! | \Psi \rangle$,
and the expectation value $P =\langle \uparrow \downarrow \uparrow \ldots \downarrow \! \!  | {\cal P} | \Psi \rangle$
where ${\cal P} = \sum_{i \in \Omega, j \in \bar{\Omega}}  S^+_i S^-_j + S^+_j S^-_i$ has a simple representation as a Matrix Product Operator, using the techniques described in Ref.~\onlinecite{McC}.
The VB entanglement entropy is then $S^{\text{VB}}=-\ln(2) P / N$.

\begin{figure*}
    \includegraphics*[width=1.35\columnwidth]{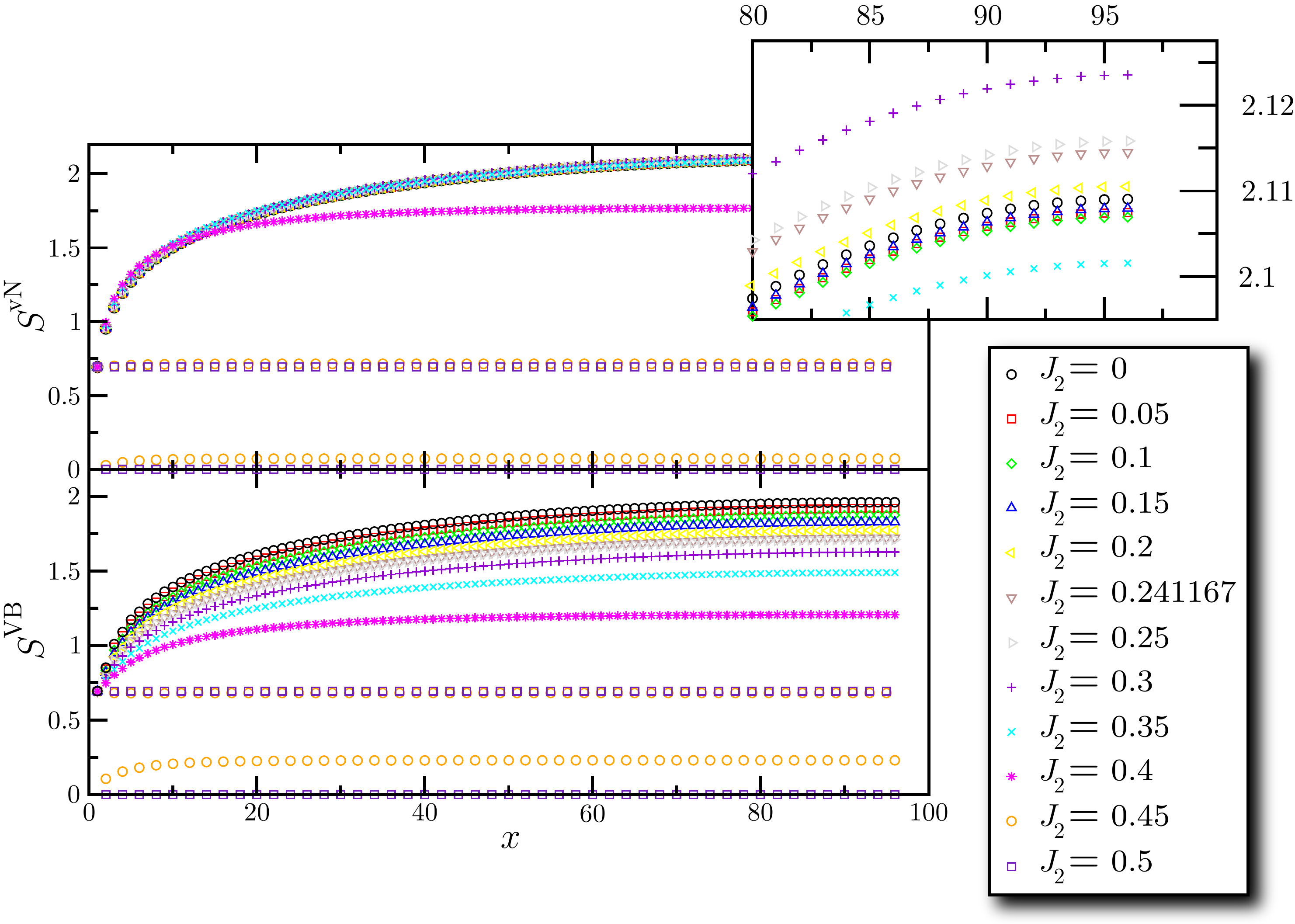}
  \caption{(Color online) Scaling of von Neumann $S^\mathrm{vN}$ (top panel) and Valence Bond  $S^\mathrm{VB}$(bottom panel) entanglement entropies as a function of block size $x$, for different values of the frustrating coupling $J_2$. Chain length is $L=192$. Top inset is a zoom of $S^\mathrm{vN}$  for $x$ close to $L/2$.}
  \label{fig:comp1}
\end{figure*}

\section{Results for the $J_1$-$J_2$ spin chain}
\label{sec:results}
\subsection{Model and simulation details}

We now present numerical results for the frustrated $J_1$-$J_2$ spin chain. $\ket{\Psi}$ in Eq.~(\ref{eq:VBEE_proj}) is taken as the ground-state of the $S=1/2$ Hamiltonian :
\begin{equation}
    H = \sum_{i=1}^L J_1 {\mathbf S}_i.{\mathbf S}_{i+1} + J_2 {\mathbf S}_i.{\mathbf S}_{i+2}
  \label{eq:hamiltonian}
\end{equation}
where we set $J_1=1$ and will vary $J_2$. The physics of this spin chain is well understood: for $J_2$ smaller than the critical value $J_2^c\simeq 0.241167$ (Ref.~\onlinecite{Eggert}), the system displays antiferromagnetic quasi-long range order, with algebraically decaying spin correlations. For $J_2>J_2^c$, the system is located in a gapped dimerized phase which spontaneously breaks translation symmetry. We will study both VB and vN EE entropies in this system in both phases. 

Results for $J_2 \leq 0.5$ were obtained with DMRG. We used samples
with $L=64$, $128$ and $192$ and periodic boundary conditions in order
to avoid dimerization effects in the entanglement
entropy~\cite{Laflorencie,Alet}, which complicate the finite-size
analysis. Up to $m=1092$ SU($2$) states (roughly corresponding to 4000
usual U($1$) states) have been kept for the largest samples. A long
warm-up procedure has been used, by performing between 10 and 50
sweeps each time $m$ was increased by 50. Convergence has been checked
by ensuring that the energy does not change significantly on more than
20 sweeps for the last value of $m$. Truncation error per site and variance per site $(H-E)^2/L$ were always at most $10^{-10}$ for the largest systems.
For these periodic boundary
conditions, a two-sites version of the DMRG algorithm has been used. We
will essentially present results for the largest $L=192$ chains, but
will occasionally show data for smaller $L$ when a discussion of
finite-size effects is necessary. 

Prior to calculating the scalar
products with the reference state, we use the Wigner-Eckart theorem to
project the $SU(2)$ groundstate to $U(1)$, thus giving direct access
to the axis-dependent spin vector operators. 

Later in the paper, we will present results for $J_2>0.5$ which were obtained with ED for chains of length up to $L=32$. For large values of $J_2$, the DMRG algorithm has more difficulties to converge, even for small samples - a fact which has already been reported~\cite{WhiteAffleck}. Also, some intrinsic difference shows up in the definition of the VB EE in this case due to the rapid vanishing of the Marshall sign in the ground-state wave function. In that situation, the analysis of the definition as well as meaning of the VB EE is different and will be discussed in Sec.~\ref{sec:marshall}.

We finally note that the vN EE of the ground-state of this spin chain was studied previously~\cite{EEJ1J2}, albeit on smaller systems, with ED techniques. As we will see later, the use of large systems is necessary to locate precisely the quantum phase transition at $J_2^c$.

\subsection {Results for $J_2\leq 0.5$}

\begin{figure*}
    \includegraphics*[width=1.4\columnwidth]{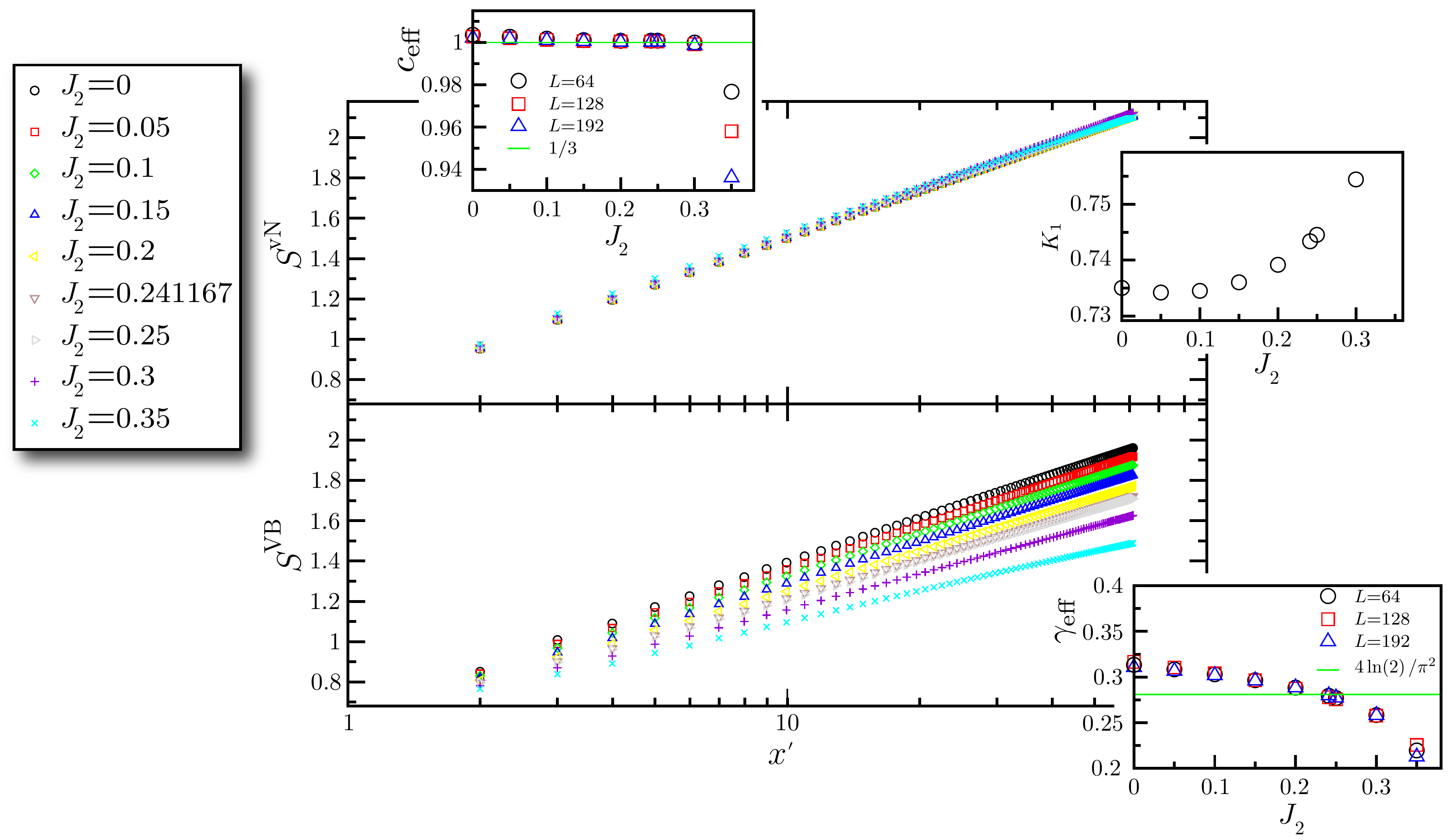}
  \caption{(Color online) Scaling of von Neumann $S^\mathrm{vN}$ (top panel) and Valence Bond  $S^\mathrm{VB}$(bottom panel) entanglement entropies as a function of conformal block size $x'=L/\pi \sin(\pi x /L)$ in a log-linear scale, for different values of the frustrating coupling $J_2$. Chain length is here $L=192$. Top left (bottom right) inset shows the value of the coefficient of a log fit for both von Neumann (Valence Bond) entanglement entropies as a function of $J_2$, for different system sizes (see text for details). Top right inset gives the value of the non-universal additive constant $K_1$ obtained from the same fit  of $S^{\mathrm vN}$, as a function of $J_2$ for $L=192$.}
  \label{fig:comp2}
\end{figure*}

We will present in this section results obtained for vN and VB entanglement entropies in parallel. We first present raw data for both entropies as a function of the block size $x$ for different values of $J_2$ in Fig.~\ref{fig:comp1}. Data are shown only for $x\in [0,L/2]$ (we have checked that curves are symmetric around $L/2$). Both sets of curves show a similar behavior: on the scale of the figure, one can distinguish between curves which converge to a constant (for $J_2\geq0.4$) and those which grow slowly but steadily with $x$. The difference between the two entropies appears on the former cases, where curves for different $J_2$ appear more shifted for $S^\mathrm{VB}$. The shift also exists for $S^\mathrm{vN}$ but is smaller (see zoom around $x\sim L/2$ in the inset of the figure). 

One should also note the clear dimerization of both entropies for large $J_2$: this is naturally expected at the Majumdar-Ghosh~\cite{MG} point $J_2=1/2$ where $S^\mathrm{VB}$ and $S^\mathrm{vN}$ are strictly equal to $0$ for even $x$ and $\ln(2)$ for odd $x$. Note that this dimerization effect comes from the intrinsic dimerized nature of the ground state in this region, and not from the boundary conditions as in Ref.~\onlinecite{Laflorencie}.

Let us concentrate now on the upper beam of curves, for $J_2\leq 0.35$. From conformal invariance of the
ground-state in the critical phase, the use of the conformal block length $x'=L/\pi \sin( \pi x
/L)$ should be useful for systems with periodic boundary conditions: in the critical phase, vN EE should
scale as $S^\mathrm{vN}=c/3 \ln(x')+K_1$ whereas $S^\mathrm{VB}=\gamma \ln(x')+K_2$.

Fig.~\ref{fig:comp2} displays both entropies versus $x'$ in a log-linear scale. All curves seem at first glance
linear with approximately the same slope, except for $J_2=0.35$ where a
crossover to a constant regime can be identified: this is well visible for $S^\mathrm{vN}$ in the figure, but is also the case for $S^\mathrm{VB}$ when zoomed in.

We fit the curves to a form $S^\mathrm{vN}=c_\mathrm{eff}/3 \ln(x') + K_1$ and $S^\mathrm{VB}=\gamma_\mathrm{eff} \ln(x') + K_2$ within the window $x'>10$.
Fits are excellent and lead to $c_\mathrm{eff}$ (respectively $\gamma_\mathrm{eff})$ very close to the CFT prediction $1$ (resp.  $4\ln(2)/\pi^2$) in the critical phase. The values of the fitted $c_\mathrm{eff}$ and $\gamma_\mathrm{eff}$ are displayed in the
left insets of Fig.~\ref{fig:comp2}, as obtained for the three different samples sizes $L$ used in this study. The finite-size effects are found to be small on both quantities. 

Several remarks are in order at this stage :

{\it (i)}  The finite size dependence of the fitted values indicate that $J_2=0.35$ is clearly not in the critical regime as we already guessed from a visual inspection of curves.   

{\it (ii)} It is quite interesting to note that both $\gamma_\mathrm{eff}$ and $c_\mathrm{eff}$ are not strictly equal to the predicted values for low $J_2$ (including $J_2=0$) but are getting closer monotonously to the theoretical predictions when increasing $J_2$. This effect can clearly be seen for $\gamma_\mathrm{eff}$, but also exist (even if small) for $c_\mathrm{eff}$. 

{\it (iii)} The values closest to $1$ and $4\ln(2)/\pi^2$ are found to be precisely at $J_2^c=0.241167$. The fitted $c_\mathrm{eff}$ is smaller than $1$ for $J_2=0.30$, and $\gamma_\mathrm{eff}$ is smaller than  $4\ln(2)/\pi^2$ for $J_2\geq 0.241167$.
  
\begin{figure*}
    \includegraphics*[width=1.4\columnwidth]{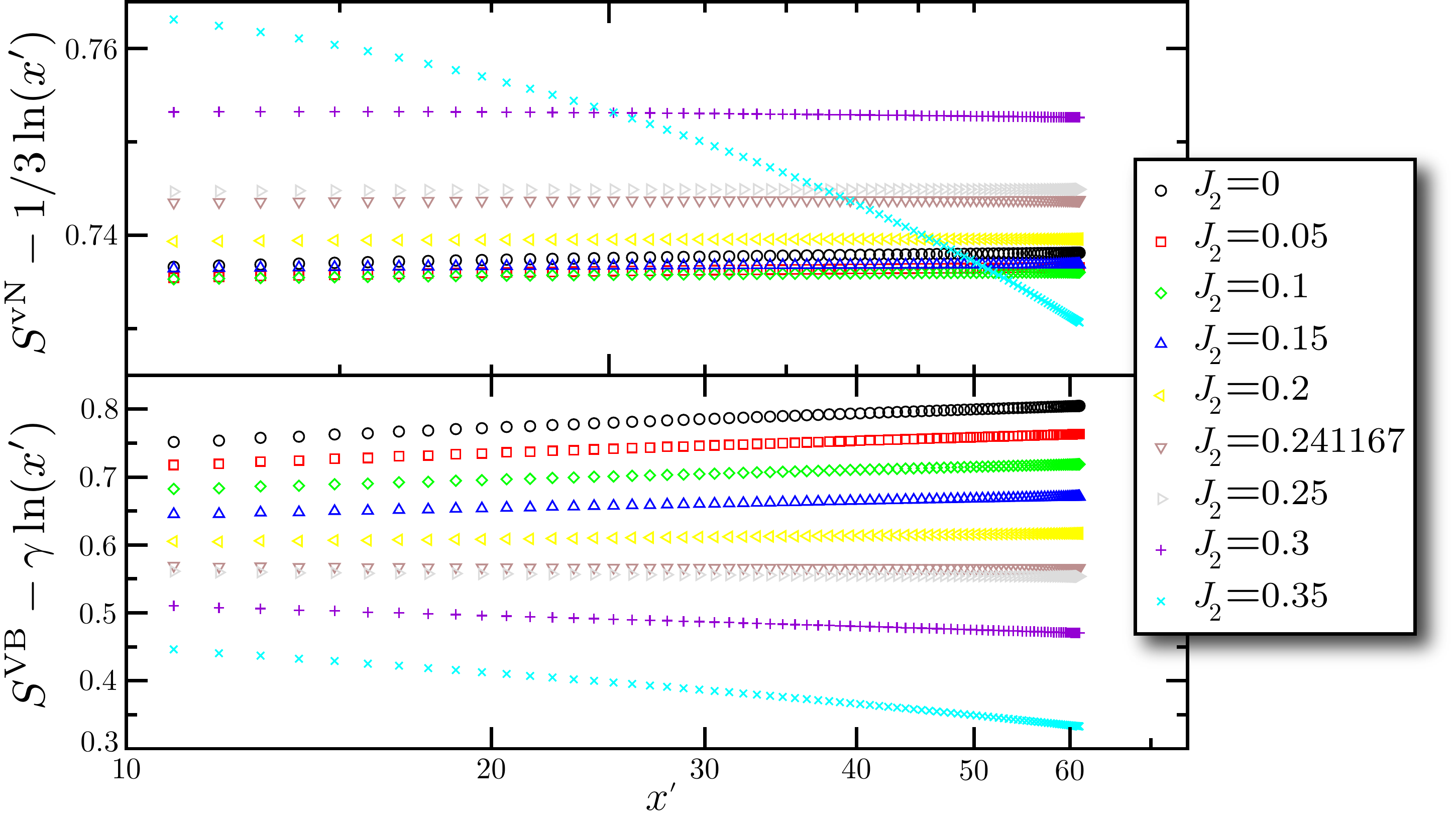}
  \caption{(Color online) Difference of the entanglement entropies to the predicted analytical values $S^\mathrm{vN}-1/3 \ln(x')$ (top panel) and $S^\mathrm{VB}-\gamma  \ln(x')$, as a function of conformal block size $x'$, for different values of the frustrating coupling $J_2$. Chain length is here $L=192$. }
  \label{fig:comp3}
\end{figure*}

The two former points lead to the following interpretation : given the existence of dangerously irrelevant operators in the critical phase (but not at $J_2^c$), we expect that they could influence the effective value of the central charge and $\gamma$ as measured from a fit of EE on finite systems (they should not in the thermodynamic limit). The strength of their influence
decreases as one approaches the critical point where they vanish. This scenario sounds plausible for $\gamma$: indeed in the field-theoretical description of the Heisenberg chain, $\gamma$ is related to the coupling constant of the free boson field whose numerical determination is known to suffer from log corrections due to dangerously irrelevant operators.  A similar effect has been recently predicted for the effective central charge in presence of marginally irrelevant operators~\cite{Cardy}, with the prediction that  $c_\mathrm{eff}<c$. Raw fits of the form $S^\mathrm{vN}=c_\mathrm{eff}/3 \ln(x') + K_1$ appear to give a value of  $c_\mathrm{eff}$ slightly larger than $1$ (in all cases less than $1 \%$). We find however that the simultaneous fits of $c_\mathrm{eff}$ and $K_1$ actually affect the determination of $c_\mathrm{eff}$. A more precise fitting procedure~\cite{note.calabrese} along the lines of Ref.~\onlinecite{Xavier} produce values of $c_\mathrm{eff}<1$ in agreement with Ref.~\onlinecite{Cardy}. Details of such a precise estimation of $c_\mathrm{eff}$ are left for a future study (we checked that a similar analysis for $\gamma_\mathrm{eff}$ does not affect the results displayed in Fig.~\ref{fig:comp2}).

The analysis above explains why first simulations of the unfrustrated Heisenberg chain (at $J_2=0$) indicate that the scaling of the VB EE was identical to the one of vN EE: indeed the fitted value of $\gamma_\mathrm{eff}\sim 0.310$ is closer to $1/3$ than to $4 \ln(2)/ \pi^2\simeq 0.281$. We note that the transfer matrix estimates of  $\gamma_\mathrm{eff}$ also display such a small discrepancy for the Heisenberg chain~\cite{Jacobsen}. The numerical results of Ref.~\onlinecite{Jacobsen} for other spin chains not suffering from these log corrections appear to be in much better agreement with the analytical predictions, confirming this scenario.

Actually, the vanishing of these log corrections appear as a way to detect on finite systems the quantum critical point $J_2^c$ through the log scaling of $S^\mathrm{VB}$ (and possibly $S^\mathrm{vN}$), as long as the exact values $\gamma=4\ln(2)/\pi^2$ and $c=1$ are known. If these values were not available, it would be more difficult to judge on the extent of the critical phase. Indeed from the sole quality of the fits, data at $J_2=0.25$ and $J_2=0.30$ (which are theoretically located in the gapped phase) are compatible with a critical scaling. This is certainly due to the small simulation length used $L$ with respect to the large correlation length close to $J_2^c$. 

Finally, we discuss the behavior of the constants $K_1$ and $K_2$.  Both constants are non-universal and are {\it a priori} not related. The fitted value of $K_1$ is shown in the right inset of Fig.~\ref{fig:comp2}, and displays a non-monotonous behavior (especially at small $J_2$). This non-monotonous behavior can also directly be seen on the raw $S^{\rm vN}$ data at $L/2$ (inset of Fig.~\ref{fig:comp1}) . On the other hand, and this can be noticed without  a fit, the constant $K_2$ for $S^\mathrm{VB}$ decreases monotonously with $J_2$.

The final transformation which summarizes these results consists in directly
subtracting the expected exact value from both entanglement entropies. Fig.~\ref{fig:comp3} displays $S^\mathrm{vN}-1/3\ln(x')$ and $S^\mathrm{VB}-\gamma \ln(x')$ versus $x'$. Curves should saturate to the constants $K_1$ and $K_2$ respectively, which they do (except obviously for $J_2=0.35$) on this scale. Zooming in, one observes that all curves for $S^\mathrm{vN}$ grow in a very smooth way, except for $J_2=0.30$ which actually decreases with $x'$ (for $S^\mathrm{VB}$, all curves for $J_2\geq 0.241167$ tend to decrease when increasing block size $x'$). This is in correspondence with the fitted $c_{\rm eff} < 1$ for this value of $J_2$. The flattest curves are observed for $J_2=0.241167\sim J_2^c$ and $J_2=0.25$ for both entropies, in agreement with our previous observations. 

\subsection {Results for $J_2>0.5$}
\label{sec:marshall}

Knowing the nodes and the signs of all coefficients of a correlated wavefunction is a difficult task. 
Indeed, such an information could allow to design a sign-free QMC algorithm as well as help on building variational wave-functions.
In this context, one can define the so-called Marshall-Peierls sign~\cite{Marshall}
\begin{equation}\label{marshall.eq}
 s (|\Psi\rangle) = \sum_i (-1)^{N_\uparrow^{\cal B}} a_i |a_i|
\end{equation}
where the sum runs over the $|\psi_i\rangle$ $\hat{S}_z$ basis states
 and with the wavefunction given by $|\Psi\rangle = \sum_i a_i |\psi_i\rangle$. $N_\uparrow^{\cal B}$ counts the number of up spins on the $\cal{B}$ sublattice so that obviously, $s$ depends on the choice of bipartition. 
 
\begin{figure}
  \begin{center}
    \includegraphics*[width=\columnwidth]{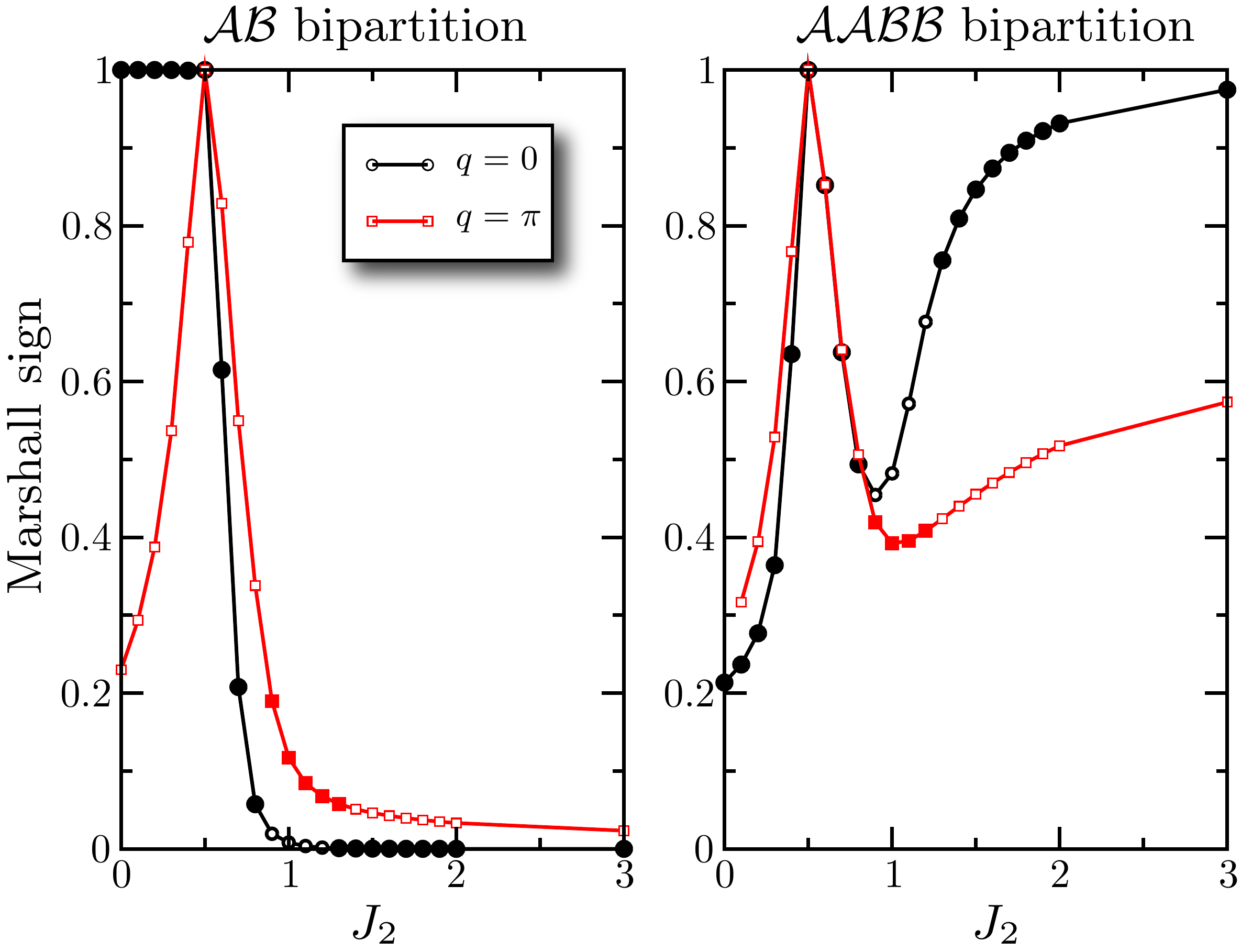}   
   \end{center}
  \caption{(Color online) Marshall sign vs $J_2$ for  a $L=32$ chain for both the lowest $q=0$ and $q=\pi$ eigenstates. Left : ${\cal AB}$ partition for the Marshall sign. Right : ${\cal AABB}$ partition. Ground-state results are denoted with filled symbols. }
  \label{fig:marshall}
\end{figure}

For non-frustrated Heisenberg model,  it can be shown that the ground-state has $s=1$ with the natural choice of bipartition~\cite{LSM}. Frustration will spoil this result, although the sign may not drop suddenly (see for instance 1d or 2d $J_1-J_2$ model~\cite{Richter1994}).  On left panel of Fig.~\ref{fig:marshall}, we present ED data for the frustrated Heisenberg chain. With the natu\-ral ${\cal AB}$ bipartition, the Marshall sign stays extremely close to 1 for $0\leq J_2 \leq 0.5$ but starts to deviate substantially beyond. Since the ground-state oscillates between having momentum 0 and $\pi$, we plot both values of $s$. 

\begin{figure}
  \begin{center}
    \includegraphics*[width=\columnwidth]{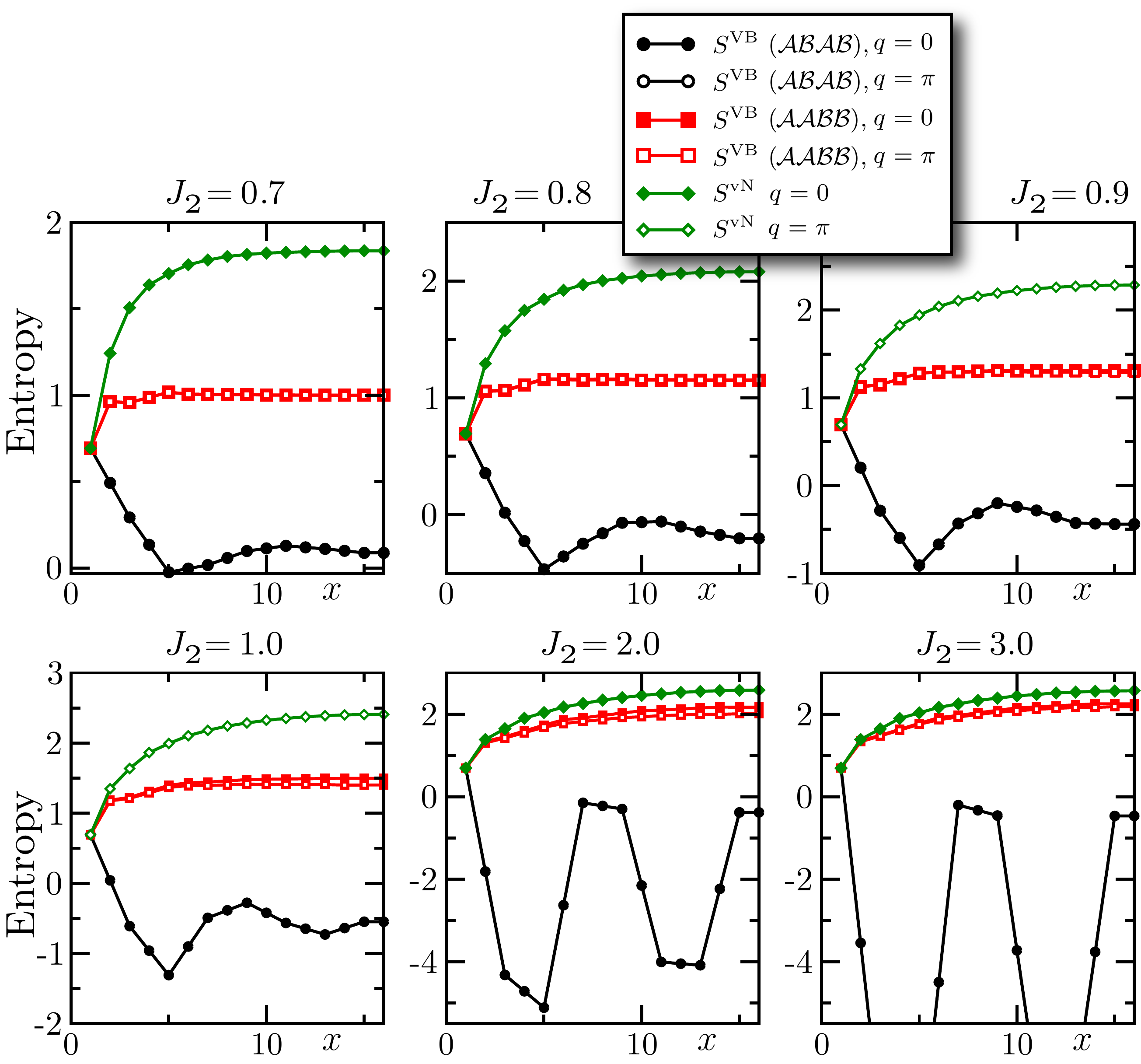}
   \end{center}
  \caption{(Color online) Entropy vs block size for various frustration parameter $J_2$ on $L=32$ system obtained with ED. 
     Several $S^\mathrm{VB}$ are plotted for the lowest $q=0$ (filled symbols) and $q=\pi$ (open) eigenstates, and for both choices of bipartition corresponding to to Ising configurations for $J_2=0$ (${\cal ABAB}$) and $J_1=0$ (${\cal AABB}$) respectively. $S^\mathrm{vN}$ is plotted for the ground-state. 
   }
  \label{fig:entropyED}
\end{figure}

Fig.~\ref{fig:entropyED} shows ED data for various entropies. In particular, since $S^\mathrm{VB}$  depends on the choice of bipartition, one may wonder what to choose. Usually, one is guided by the Ising solution: for small $J_2/J_1$, it is natural to choose a ${\cal ABAB} \ldots$ bipartition, while for large $J_2/J_1$, the system will behave as two decoupled Heisenberg chains with twice as large lattice spacing, meaning that bipartition should be of the form ${\cal AABB} \ldots$ The Marshall sign for this ${\cal AABB}$ partition is presented in the right panel of Fig.~\ref{fig:marshall}. 

Of course, the intermediate region with maximal frustration has no preferred bipartition. Moreover, since the ground-state oscillates between $q=0$ and $q=\pi$, we plot both entropies. Still, as can be seen from its definition Eq.~(\ref{eq:VBEE_proj}),  $S^\mathrm{VB}$  is only defined when the overlap between the ground-state and the classical N\'eel state is finite. Unfortunately, the ${\cal ABAB}$ N\'eel state has no projection in the singlet $q=\pi$ sector as it is invariant by a combination of lattice translation and spin reversal. However, for ${\cal AABB}$ bipartition, we can compute $S^\mathrm{VB}$  for both lowest $q=0$ and $q=\pi$ states and it turns out that data are very similar (sometimes they cannot be distinguished on the scale of Fig.~\ref{fig:entropyED}), although states are quite different (see their Marshall sign in Fig.~\ref{fig:marshall}). 

By comparing Fig.~\ref{fig:entropyED} and Fig.~\ref{fig:marshall}, we observe that when the Marshall sign is too small, $S^\mathrm{VB}$ has absolutely no meaning (and can even be negative). On the other hand, is $s$ is large enough, or said differently, if we choose the bipartition that maximizes $s$, then $S^\mathrm{VB}$ behaves much better and follows the same trend as $S^\mathrm{vN}$, that is both converge to a constant for large enough block size. In fact, for large $J_2/J_1$ where the Marshall sign becomes again close to 1 for the ${\cal AABB}$ choice, we observe that both entropies become more and more similar. 

\section{Discussion}
\label{sec:disc}
 
In this paper, we studied the behavior of the Valence Bond
Entanglement Entropy of the frustrated $J_1$-$J_2$ spin chain, and
offered a direct comparison with the von Neumann entropy. Numerical
DMRG calculations indicate that both entropies scale logarithmically
with block size in the critical phase, and converge in the gapped
dimerized phase of this model. The study of the VB EE has been made
possible in this frustrated model through a formulation of
valence bond occupation number, which extends this notion out of the
Valence Bond basis.
 
We now discuss several interests of studying entanglement in quantum spin systems through a valence bond measure, and point out some open issues.

First, as based on the example of the $J_1$-$J_2$ model, the scaling of $S^\mathrm{VB}$ with the block size allows to differentiate critical from gapped phases in one dimension, similarly to $S^\mathrm{vN}$. Moreover, the knowledge of the exact prefactor~\cite{Jacobsen} of the log scaling permits a relatively precise determination of the quantum critical point at $J_2^c$ with finite-size data (this is however due to the vanishing of log corrections at this particular point, a non-generic feature). Note that this knowledge can be useful as the scaling of  $S^\mathrm{VB}$ might now be used to characterize uniquely the unknown phase of a new model. $S^\mathrm{VB}$ can be computed for the Q-states Potts model using the loop language of Ref.~\onlinecite{Jacobsen} and there, the prefactor of the logarithmic scaling depends on Q and is therefore indeed different for the different critical points encountered in the Potts model. This is similar to the scaling of $S^\mathrm{vN}$ which allows the determination of the central charge - the knowledge of which entirely determines the CFT at play for minimal models. 

It is possible to compute $S^\mathrm{VB}$ directly in the thermodynamic limit using the infinite-size algorithm iDMRG~\cite{idmrg} or iTEBD~\cite{itebd}. In this formulation, the number of basis states in the calculation controls the spectrum transfer matrix of the system, which gives a scaling of the correlation length $\xi \propto m^\kappa$ at criticality, where $\kappa$ is a function of the central charge~\cite{pollmann}. However, the effective boundary condition for the transfer matrix, and hence the form of the corrections to scaling of the entropies $S^\mathrm{VB}$ and $S^\mathrm{vN}$, will be different to the case of periodic boundary conditions, and we leave this analysis for a future study. Besides, knowing the corrections to scaling induced on a finite system by marginally irrelevant operators is interesting by itself (see for instance Ref.~\onlinecite{Cardy}).  

In dimension higher than 1, it was already demonstrated~\cite{Alet, Chhajlany,Kallin} that the scaling of $S^\mathrm{VB}$ discriminates between gapped and gapless phases. Since numerical calculations are possible in $d>1$ with QMC VB methods~\cite{Sandvik}, it would be of high interest to perform a systematic study of the scaling of $S^\mathrm{VB}$ in different phases of quantum spin models. Several questions are in order: for instance, do the multiplicative log corrections observed for the 2d Heisenberg model have a physical interpretation? Are prefactors of the scaling of $S^\mathrm{VB}$ universal within a phase or at a quantum critical point~\cite{Chhajlany}, as observed in 1d? 

We note that the techniques described here for calculating the VB entanglement entropy can easily be applied to higher dimensional tensor network algorithms such as PEPS~\cite{Verstraete} where the necessary scalar products are similarly easily computed. This opens the door to studies of the VB entanglement entropy for frustrated 2D systems~\cite{Frustrated2d}. 

Our study on frustrated systems also sheds lights on the importance of a good physical choice for the bipartition used in the definition of the VB EE, and its relation to the existence of a Marshall sign rule (or a large Marshall sign) in the wave-function under study. When the Marshall sign is exactly or close to $1$, the resulting choice of bipartition (or equivalently reference state $\vert R_S \rangle$ in Eq.~(\ref{eq:projection_mixed_normalized})) produces a Valence Bond entanglement entropy that closely follows the scaling of the von Neumann entropy. This suggests another route to quantify entanglement in a wave-function through its projection over a well-chosen (physical) reference state.
 
Another interesting situation which we have not discussed in this work is the one of strongly disordered spin systems, where $S^\mathrm{VB}$ and $S^\mathrm{vN}$ coincide~\cite{Refael} (after averaging over disorder). This is the case in the random singlet phase, where the low-energy physics is dominated by a single valence-bond state, as remarked in Ref.~\onlinecite{Alet} and more recently by Tran and Bonesteel~\cite{Tran}. As pointed out by these authors, the study of the {\it fluctuations} of the number of VBs crossing the boundary provide additional physical insights in this situation. 

Finally, we comment on the usefulness of the formulation Eq.~(\ref{eq:projection_mixed_normalized}) of the valence bond occupation number. It clearly points towards generalizations of $S^\mathrm{VB}$ for situations not explored before, for instance for spins higher than $1/2$, as well as for systems which lack SU($2$) symmetry. In the latter case, the direct interpretation in terms of $SU(2)$ VBs is not possible anymore and the physical meaning of $n_{(i,j)}$ has first to be clarified. It would be interesting for instance to look for the relation to q-deformed singlets~\cite{Pasquier}, which are used in Ref.~\onlinecite{Jacobsen} to extend the VBEE to Potts models. Another high interest of Eq.~(\ref{eq:projection_mixed_normalized}) is that it allows analytical insights on the distribution of valence bonds and their correlations in a singlet ground-state~\cite{Oshikawa}.



\begin{acknowledgments}
We warmly thank O. Giraud for a remark that led to this work, M. Oshikawa for collaboration on related work and P. Calabrese and F. Becca for very useful discussions. We thank CALMIP for allocation of CPU time. This work is supported by the French ANR program ANR-08-JCJC-0056-01.
\end{acknowledgments}



\begin{thebibliography}{99}



\bibitem{review} P. Calabrese, J. Cardy, and B. Doyon, J. Phys. A {\bf 42}, 500301 (2009).

\bibitem{area} M. Srednicki, Phys. Rev. Lett. {\bf 71}, 666 (1993) ; J. Eisert, M. Cramer and M.B. Plenio, Rev. Mod. Phys. {\bf 82}, 277 (2010).

\bibitem{Calabrese} C. Holzhey, F. Larsen, and F. Wilczek, Nucl.Phys.B {\bf 424}, 443 (1994); G. Vidal, J.I. Latorre, E. Rico and A. Kitaev, Phys. Rev. Lett. {\bf 90}, 227902 (2003); J.I. Latorre, E. Rico and G. Vidal, Quant.Inf.Comput. {\bf 4}, 48 (2004); P. Calabrese and J. Cardy, J. Stat. Mech. P06002 (2004).

\bibitem{Fradkin} E. Fradkin and J.E. Moore, Phys. Rev. Lett. {\bf 97}, 050404 (2006).

\bibitem{Stephan} J.-M. St\'ephan, S. Furukawa, G. Misguich and V. Pasquier, Phys. Rev. B {\bf  80}, 184421 (2009).

\bibitem{bosons1} M. Cramer, J. Eisert, M. B. Plenio and J. Dreissig, Phys. Rev. A {\bf 73}, 012309 (2006).

\bibitem{bosons2} T. Barthel, M.-C. Chung and U. Schollw\"ock, Phys. Rev. A  {\bf 74}, 022329 (2006).

\bibitem{fermions} M.M. Wolf, Phys. Rev. Lett. {\bf 96}, 010404 (2006); D. Gioev and I. Klich, {\it ibid} {\bf 96}, 100503 (2006).

\bibitem{White} S. R. White, Phys. Rev. Lett. {\bf 69}, 2863 (1992).

\bibitem{Buividovich} P.V. Buividovich and M.I. Polikarpov, Nucl. Phys. B {\bf 802}, 458 (2008).

\bibitem{Caraglio} M. Caraglio and F. Gliozzi, JHEP {\bf 11} (2008) 076.

\bibitem{Alet}  F. Alet, S. Capponi, N. Laflorencie and M. Mambrini, Phys. Rev. Lett. {\bf 99}, 117204 (2007).

\bibitem{Chhajlany} R.W. Chhajlany, P. Tomczak and A. W\'ojcik, Phys. Rev. Lett. {\bf 99}, 167204 (2007).

\bibitem{Sandvik} A.W. Sandvik, Phys. Rev. Lett. {\bf  95}, 207203 (2005); A.W. Sandvik and H.-G. Evertz, Phys. Rev. B {\bf 82}, 024407 (2010). 

\bibitem{Jacobsen} J.L. Jacobsen and H. Saleur, Phys. Rev. Lett. {\bf 100}, 087205 (2008). 

\bibitem{Kallin} A.B. Kallin, I. Gonz\'alez, M.B. Hastings and R.G. Melko, Phys. Rev. Lett. {\bf 103}, 117203 (2009).

\bibitem{Hastings} M.B. Hastings, I. Gonz\'alez, A.B. Kallin and R.G. Melko, Phys. Rev. Lett. {\bf 104}, 157201 (2010).

\bibitem{Stoch} F.C. Alcaraz, V. Rittenberg and G. Sierra, Phys. Rev, E {\bf 80},  030102(R) (2009); F.C. Alcaraz and V. Rittenberg,  J. Stat. Mech. P03024 (2010); P. Calabrese J. Stat. Mech. N05001 (2010) ; A.L. Owczarek, J. Stat. Mech. P12004 (2009).

\bibitem{Mambrini} M. Mambrini, Phys. Rev. B {\bf 77}, 134430 (2008).

\bibitem{Rumer} G.~Rumer, E.~Teller and H.~Weyl~: Nachr. Gott., Math-physik. Klasse, 499 (1932). See also H.~N.~Temperley and E.~H.~Lieb, Proc. Roy. Soc. Lond. A. {\bf 322}, 251-280 (1971) and R.~Saito J. Phys. Soc. Japan {\bf 59}, 482-491 (1990).

\bibitem{Beach} K. S. D. Beach, F. Alet, M. Mambrini and S. Capponi, Phys. Rev. B {\bf 80}, 184401 (2009).

\bibitem{Ref_Lanczos} See for instance N. Laflorencie and D. Poilblanc, 
 Lect. Notes Phys. {\bf 645}, 227 (2004) and references therein. 

\bibitem{McC} I.P. McCulloch, J. Stat. Mech. P10014 (2007).

\bibitem{EEJ1J2} R.W. Chhajlany, P. Tomczak, A. W\'ojcik, and J. Richter, Phys. Rev. A {\bf 75}, 032340 (2007).

\bibitem{Marshall}  W. Marshall, Proc. R. Soc. London Ser. {\bf A 232}, 48 (1955); E.H. Lieb and D.C. Mattis, J. Math. Phys. {\bf 3},
749 (1962).

\bibitem{Eggert} S. Eggert, Phys. Rev. B. {\bf 54}, R9612 (1996).

\bibitem{Laflorencie} N. Laflorencie, E.S. S\o rensen, M.-S. Chang and I. Affleck, Phys. Rev. Lett. {\bf 96}, 100603 (2006).

\bibitem{WhiteAffleck} S.R. White and I. Affleck, Phys. Rev. B {\bf 54}, 9862 (1996).

\bibitem{MG} C.K. Majumdar and D.K. Ghosh, J. Math. Phys. {\bf 10}, 1388 (1969); {\it ibid} 1399 (1969).

\bibitem{Cardy} J. Cardy and P. Calabrese, J. Stat. Mech. P04023 (2010) .

\bibitem{Xavier}  J.C. Xavier, Phys. Rev. B {\bf 81}, 224404 (2010).

\bibitem{note.calabrese} P. Calabrese, private communication.

\bibitem{LSM} E.H. Lieb, T.D. Schultz, and D.C. Mattis, Ann. Phys. (NY) {\bf 16}, 407 (1961).

\bibitem{Richter1994} J. Richter, N.B. Ivanov, and K. Retzlaff, Europhys. Lett. {\bf 25}, 545 (1994).

\bibitem{idmrg} I. P. McCulloch, arxiv:0804.2509 (unpublished).

\bibitem{itebd} G. Vidal, Phys. Rev. Lett. {\bf 98}, 070201 (2007).

\bibitem{pollmann} F. Pollmann, S. Mukerjee, A. M. Turner and J. Moore, Phys. Rev. Lett. {\bf 102}, 255701 (2009).

\bibitem{Verstraete} F. Verstraete and J.I. Cirac, cond-mat/0407066; F. Verstraete, M. M. Wolf, D. Perez-Garcia, and J. I. Cirac,  Phys. Rev. Lett. {\bf 96}, 220601 (2006).

\bibitem{Frustrated2d} V. Murg, F. Verstraete and J. I. Cirac, Phys. Rev. B {\bf 79}, 195119 (2009)

\bibitem{Refael} G. Refael and J.E. Moore, Phys. Rev. Lett. {\bf 93}, 260602 (2004).

\bibitem{Tran} H. Tran and N.E. Bonesteel, arxiv:0909.0038 (unpublished).

\bibitem{Pasquier} V. Pasquier and H. Saleur, Nucl. Phys. B {\bf 330}, 523 (1990).

\bibitem{Oshikawa} M. Oshikawa, D. Schwandt, and F. Alet, unpublished.




\end{thebibliography}
\end{document}